\documentclass[useAMS,usenatbib]{mn2e}
\usepackage{times}
\usepackage{graphics,epsfig}
\usepackage{graphicx}
\usepackage{amssymb}
\title[{\rm H}~{\sc i} structure associated with V458~Vul]{An H~{\sc i} shell-like structure associated with nova V458 Vulpeculae?} 
\author[Roy et al.]{Nirupam Roy$^{1}$\thanks{Contact author's E-mail: nroy@aoc.nrao.edu (NR); NR is a Jansky Fellow of the National Radio Astronomy Observatory (NRAO). The NRAO is a facility of the National Science Foundation operated under cooperative agreement by Associated Universities, Inc.}, 
N. G. Kantharia$^{2}$, S. P. S. Eyres$^{3}$, G. C. Anupama$^{4}$, M. F. Bode$^{5}$, 
\newauthor T. P. Prabhu$^{4}$ and T. J. O'Brien$^{6}$\\
     $^{1}$National Radio Astronomy Observatory, 1003 Lopezville Road, Socorro, NM 87801, USA\\
     $^{2}$National Centre for Radio Astrophysics, TIFR, Post Bag 3, Ganeshkhind, Pune 411 007, India\\
     $^{3}$Jeremiah Horrocks Institute, University of Central Lancashire, Preston PR1 2HE, UK\\
     $^{4}$Indian Institute of Astrophysics, Sarjapur Road, Koramangala, Bangalore 560034, India\\
     $^{5}$Astrophysics Research Institute, Liverpool John Moores University, Birkenhead CH41 1LD, UK\\
     $^{6}$Jodrell Bank Centre for Astrophysics, University of Manchester, Manchester M13 9PL, UK}

\begin{document}
\date{Accepted yyyy month dd. Received yyyy month dd; in original form yyyy 
month dd}

\pagerange{\pageref{firstpage}--\pageref{lastpage}} \pubyear{2012}

\maketitle

\label{firstpage}

\begin{abstract}
We report the radio detection of a shell-like H~{\sc i} structure in proximity 
to, and probably associated with, the nova V458~Vul. High spectral resolution 
observation with the Giant Metrewave Radio Telescope has made it possible to 
study the detailed kinematics of this broken and expanding shell. Unlike the 
diffuse Galactic H~{\sc i} emission, this is a single velocity component 
emission with significant clumping at $\sim 0.5\arcmin$ scales. The observed 
narrow line width of $\sim 5$ km~s$^{-1}$ suggests that the shell consists of 
mostly cold gas. Assuming a distance of 13 kpc to the system, as quoted in the 
literature, the estimated H~{\sc i} mass of the nebula is about 25 M$_\odot$. 
However, there are some indications that the system is closer than 13 kpc. If 
there is a physical association of the H~{\sc i} structure and the nova 
system, the asymmetric morphology and the off-centred stellar system indicates 
past strong interaction of the mass loss in the asymptotic giant branch phase 
with the surrounding interstellar medium. So far, this is the second example, 
after GK~Per, of a large H~{\sc i} structure associated with a classical nova.
\end{abstract}


\section{Introduction}
\label{sec:intr}

Classical novae (CNe) are interacting binaries, with orbital periods of hours 
to around one day. A white dwarf (WD) accretes gas from a main sequence 
companion, and the build up leads to a thermonuclear runaway (TNR) in the 
surface material. This generates a fireball, leading to the visual brightening 
that allows us to detect these events. This is accompanied by the ejection of 
around $10^{-4}$ M$_\odot$ of material at velocities of 200 to 5000 
km~s$^{-1}$ \citep[see e.g.][for a recent review]{war08}. These episodic 
stellar explosions are commonly detected at optical wavelengths and provide 
valuable understanding of the nuclear burning and accretion phenomena.

Most of the CNe follow a maximum-magnitude-rate-of-decline (MMRD) 
relationship, where the time to decay by 2 or 3 magnitudes from optical 
maximum ($t_2$, $t_3$ respectively) is a reasonable indicator of the absolute 
magnitude \citep[e.g.][]{dow00}. As well as providing a distance estimate, 
this indicates the the underlying mechanism is fairly uniform across different 
examples. The rate of decline also defines a speed class -- from very fast to 
slow -- and it is evident that the fastest novae are also the most energetic 
(being intrinsically brightest and having the fastest ejecta). This suggests a 
link between the development of the ejecta, the progress of the explosion and 
the energetics (including any dependence on dredged up material) of the TNR 
\citep[e.g.][]{sha80}.

V458~Vul was identified as a CN on 2007 August 8 (IAU Circular 8861) with 
magnitude 9.5 \citep{nak07}. After reaching a maximum brightness of $V =8.1$, 
the optical lightcurve showed unusual multiple peaks in the following weeks. 
If these additional peaks are ignored, it seems likely that for V458~Vul $t_3 
\sim 21$ days, making this a fast nova. The MMRD relationship gives a distance 
of 13.5, 11.6 and 10.0 kpc to V458~Vul using $t_2$, $t_3$ and $t_{15}$ 
respectively \citep{dow00,wes08}. \citet{hen07} identified the star 
USNO-B1.0~$1108-0460444$ as the ``viable progenitor for the nova'' based on 
its colour and proximity to the nova position \citep[within $0\arcsec.6$;][]{nak07}. Within $5\arcsec$ of the nova position, there is no other 
source in the USNO catalogue (completeness down to $V=21$). Earlier spectra 
indicated V458~Vul to be an Fe~{\sc ii} type nova, whilst later spectra were 
consistent with an He/N type. This suggests that, in the spectral 
classification scheme of \citet{wil94}, V458~Vul is a hybrid nova 
\citep{pog08}. X-ray observations initially showed emission consistent with an 
expanding shock \citep{tsu09}, before a soft component, reminiscent of RS~Oph, 
came to dominate \citep{dra08} with a 12 hour oscillation becoming apparent, 
as this emission declined over 2008 (Swift nova consortium). In RS~Oph this 
was associated with the unveiling of continued hydrogen burning on the WD 
surface \citep{osb11}. Follow up spectroscopy has established the the orbital 
period of the binary system to be $98$ min \citep{rod10}. 

\begin{figure}
\begin{center}
\includegraphics[scale=0.42, angle=0.0]{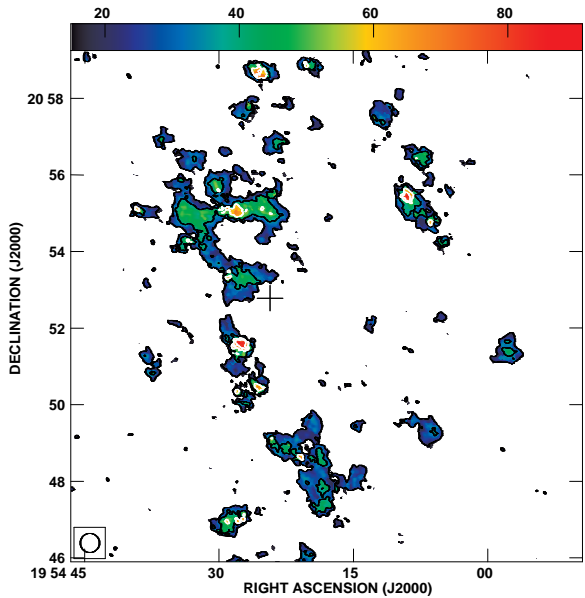}
\caption{\label{fig:fig-1} Integrated H~{\sc i} column density map of the field showing large scale H~{\sc i} emission for the velocity range of $V_{LSR} \approx -60$ to $-50$ km~s$^{-1}$. The synthesized beam size is $30\arcsec \times 30\arcsec$, and the colour scale is in mJy~km~s$^{-1}$ per beam. The contour levels are for H~{\sc i} column density of $(1,2,3,4,5)$ times $2.5 \times 10^{19}$ cm$^{-2}$. The lowest contour corresponds to a $3\sigma$ cutoff based on rms noise per channel (see \S\ref{sec:oar} for details). The optical position of V458~Vul is marked with a $+$. Note the broken shell-like structure north-east of the nova position.}
\end{center}
\end{figure}

\begin{table}
 \caption{Summary of the GMRT observation of V458~Vul}
\begin{center}
 \begin{tabular}{lc}
 \hline
Source position      & R.A. $19$h$54$m$24$s$.61$ \\
~~~~~~~(J2000)       & Dec. $+20^\circ52\arcmin52\arcsec.6$\\
GMRT pointing        & R.A. $19$h$54$m$24$s$.3$ \\
centre (J2000)       & Dec. $+20^\circ52\arcmin47\arcsec.0$\\
Date of obs.         & June 11-12, 2009              \\
Bandwidth            & $2.0$ MHz, $\sim 420$ km~s$^{-1}$ \\
No. of channels      &    $256$                      \\
Spectral resolution  & $\sim 1.6$ km~s$^{-1}$        \\
Calibrators          & 3C~48, 3C~286, 1924$+$334 \\
Observation time     &  Total: $\sim 10$ hr, On source: $\sim 4$ hr \\
\hline
\end{tabular}
\end{center}
\label{table:obs0}
\end{table}

\begin{figure}
\begin{center}
\includegraphics[scale=0.42, angle=-90.0]{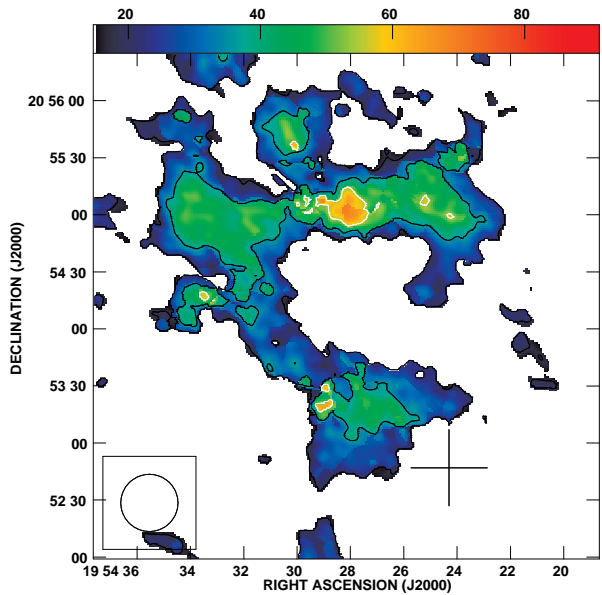}
\caption{\label{fig:fig-3} Zoom-in from Figure~\ref{fig:fig-1}, showing the H~{\sc i} column density map of the broken shell. Contour levels and gray scales are the same as in Figure~\ref{fig:fig-1}.}
\end{center}
\end{figure}

The field containing V458~Vul was observed both before and after the nova 
event, as part of the IPHAS Galactic Plane Survey. \citet{wes08} identified a 
planetary nebula in the H$\alpha$ images, with a semi-major axis of 
13\arcsec.5, with a position angle of $\sim 30\deg$, centred to within $1\arcsec$ on the nova seen in the later IPHAS images. Further follow-up imaging 
and spectroscopic observations of the nebula showed an inner nebular knot, 
brightened by the flash ionization. This nebula, with an ionized mass of $0.2$ 
M$_\odot$, is suggested to be a 14000 years old planetary nebula originating 
from the common envelope phase of the binary system \citep{wes08}. 

In this paper, we report results of our radio observation at 1420 MHz of the 
H~{\sc i} emission towards V458~Vul. This observation reveals a shell-like 
H~{\sc i} structure close to, and probably associated with, the nova. The 
observation, analysis procedure and the results are presented in Section 
\S\ref{sec:oar}. Section \S\ref{sec:dis} contains discussion on possible 
implications of these results, and our conclusions are presented in Section 
\S\ref{sec:con}.

\section{Observation, Analysis and Results}
\label{sec:oar}

\begin{figure*}
\begin{center}
\includegraphics[scale=0.28, angle=-90.0]{fig3.ps}\includegraphics[scale=0.28, angle=-90.0]{fig5.ps}\includegraphics[height=6.75cm, width=5.5cm, angle=-90.0]{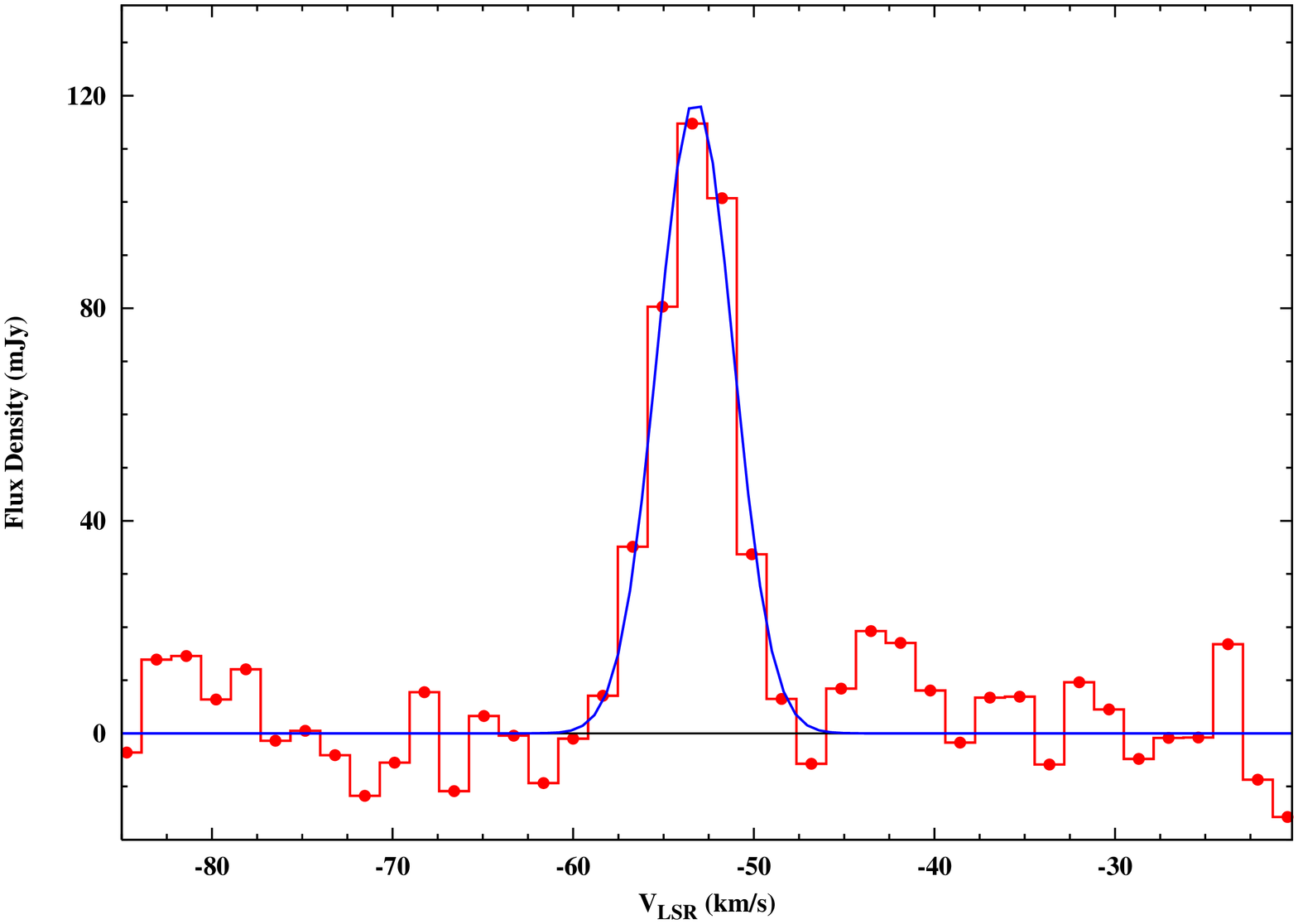}
\caption{\label{fig:fig-4} {(a):} H~{\sc i} velocity field of the region shown in Figure~\ref{fig:fig-3}. {(b):} H~{\sc i} emission near V458~Vul in different spectral channels. The LSR velocities for each channel are listed in the top right corner of the respective panels. {(c):} Integrated H~{\sc i} emission spectrum of the broken shell. The blue line is the best fit Gaussian profile with FWHM of 4.9 km~s$^{-1}$.}
\end{center}
\end{figure*}

V458~Vul was observed with the Giant Metrewave Radio Telescope (GMRT) with the 
aim of detecting any associated H~{\sc i} 21 cm emission, and to search for 
possible radio continuum from the nova shell. Bandpass calibration for this 
observation was carried out with the flux and phase calibrators, but with 
frequency-switching at the first local oscillator, using a frequency throw of 
$\pm 5$ MHz. For details of GMRT observations using frequency switching, see 
\citet{roy07}. Standard flux calibrators 3C~286 and 3C~48 (at start and end of 
the observation, respectively) were observed for $\sim 30$ minutes ($10$ 
minutes for each of the three frequency settings). A phase calibrator, 
1924$+$334, was also observed in between the target source scans with each of 
the three frequency settings. A summary of the observations is given in 
Table~\ref{table:obs0}. 

All data were analysed in the Astronomical Image Processing System of the 
National Radio Astronomy Observatory (NRAO {\small AIPS}), using standard 
procedures. Some of the data were flagged to remove interference and 
instrumental misbehaviour. After removing bad data, the standard calibration 
procedure was followed to obtain the antenna-based complex gains. The bandpass 
solution was derived from the $\pm 5$ MHz frequency setting scans of the 
calibrators, and the calibration was applied to the central frequency target 
source scans using interpolation between bandpass solutions. Next, the 
line-free channels were used to make a continuum image of the field. No radio 
continuum was detected at the nova position with a $3\sigma$ limit of $1$ mJy.

The continuum subtracted residual data were imaged with CLEANing in all 
channels. To recover weak and diffuse emission structures, data from longer 
baselines were given a lower weight while making images. The restoring beam of 
the final image cube was $30\arcsec \times 30\arcsec$ and the rms noise per 
channel was $\sim 1.3$ mJy~beam$^{-1}$. From this image cube, the moment maps 
were made using the channels with significant emission above $3\sigma$ noise 
level. Finally, this was converted to H~{\sc i} column density units assuming 
optically thin emission \citep{kul88}. As shown in Figure~\ref{fig:fig-1}, we 
have detected discrete H~{\sc i} clouds in the vicinity of V458~Vul with 
velocities $\sim -55$ km~s$^{-1}$. If the gas velocity is due to Galactic 
rotation only, the kinematic distance of the structure is $13.5 - 14.5$ kpc. A 
broken clumpy shell-like H~{\sc i} structure north-east of the nova position 
is clearly visible in this image. No such clumpiness at similar scales was 
seen in the H~{\sc i} maps at the positive velocities, indicating that the 
H~{\sc i} emission component is smoothly distributed over large angular 
scales, which is typical of the hot atomic gas detected in the H~{\sc i} 21 cm 
emission line in our Galaxy. The details of column density and velocity field 
of the broken H~{\sc i} shell are shown in Figure~\ref{fig:fig-3} and 
Figure~\ref{fig:fig-4}(a) respectively. Figure~\ref{fig:fig-4}(b) shows the 
H~{\sc i} emission in different channels for the velocity range of $-46.8$ to 
$-60.0$ km~s$^{-1}$, covering the spectral region of interest. The central 
channels show a line of sight velocity difference of $\sim 3.5$ km~s$^{-1}$ 
between the south-east and north-west rim.

The integrated spectrum of the H~{\sc i} shell is shown in 
Figure~\ref{fig:fig-4}(c). The blue line is the best fit Gaussian profile with 
a full width at half maximum (FWHM) of only $4.9 \pm 0.3$ km~s$^{-1}$. If 
there was no non-thermal broadening, this would correspond to a temperature 
$\sim 525$ K. However, as noted above, at least $\sim 3.5$ km~s$^{-1}$ is due 
to the internal dynamics of the H~{\sc i} structure. There may also be some 
contribution from turbulence. So the actual kinetic temperature is likely to 
be considerably lower. We also used the line profile to estimate an H~{\sc i} 
mass of $(25.0 \pm 2.3) \times d_{13}^2$ M$_\odot$, where $d_{13}$ is the 
distance in units of 13 kpc.

The H~{\sc i} emission towards V458~Vul (in the first Galactic quadrant) 
detected near $V_{\rm LSR} = -53$ km~s$^{-1}$ is a single narrow emission 
component while rest of the H~{\sc i} towards V458~Vul (and also towards the 
nearby phase calibrator 1924$+$334) at positive velocities has multiple 
components spread over $> 50$ km~s$^{-1}$. The H~{\sc i} emission near $-53$ 
km~s$^{-1}$ shows clumping at $\sim 0.5\arcmin$ scales. In contrast, the 
H~{\sc i} emission detected at positive velocities is more widespread and of 
relatively uniform brightness over the GMRT primary beam with no significant 
clumping below $\sim 10\arcmin$. The narrow H~{\sc i} line width of the shell 
suggests that the gas is cold. However, the estimated size and mass ($\sim 
7.6\times d_{13}$ pc and $25\times d_{13}^2$ M$_\odot$ respectively) are 
larger than the typical size and mass scales ($1-2$ pc and $2-5$ M$_\odot$ 
respectively) of the Galactic cold neutral medium. These characteristics 
indicate a possible peculiar origin (e.g. association with V458~Vul, shaped 
by an early phase of mass loss and interaction) of this shell.
 
\section{Discussion}
\label{sec:dis}

In this paper we report the discovery of a large asymmetric shell-like 
structure in H~{\sc i} to the north-east of V458~Vul. At the maximum likely 
distance of 13 kpc, the average projected linear diameter of the H~{\sc i} 
structure is about 7.6 pc. For comparison, GK~Per has a planetary nebula of 
size 6 pc with associated H~{\sc i} structure \citep{sea89,anu05}. Moreover, 
there have been reports of planetary nebulae observed in dust emission at 
infrared wavelengths which are several parsecs in size \citep{wei04}. So, a 
size of a few parsecs is not inconsistent with our interpretation of this 
structure as being associated with V458 Vul. V458~Vul is thus the second 
classical nova found to be associated with a large H~{\sc i} structure.

If the H~{\sc i} structure and the nova have a physical association, we may 
interpret this shell as a remnant of the interaction with the surrounding 
interstellar medium (ISM) in the asymptotic giant branch (AGB) phase of 
evolution of the white dwarf. The H~{\sc i} structure, which appears to be ram 
pressure swept gas, has a size of about $2\arcmin.4\times1\arcmin.8$ at a 
position angle of $\sim 30\deg$ and a systemic velocity of $-53.2$ km~s$^{-1}$. 
Interestingly, the position angle is very close to the orientation of the 
planetary nebula detected by \citet{wes08}, though the angular size of the 
planetary nebula is much smaller ($\sim 27\arcsec$). This, we believe, 
supports a common origin to this axis. It might indicate the ISM density 
structure or the axis of mass loss in the binary system. 

\subsection{The asymmetric H~{\sc i} shell: evolution in the AGB phase}

The asymmetric nature of the structure and the off-centred stellar system may 
be attributed to a number of causes including the effect of the interaction of 
the stellar wind with the ISM, the effect of the binary stellar system and 
magnetic field effects \citep[e.g.][]{kwo82,sok89,cha94,dwa96,bla01,sok01}.  

Simulations by \citet{bor90}, \citet{sok91} and more recently by \citet{war07} 
have examined the interaction of ambient ISM with systems with continuous or 
episodic mass loss. Though these are mainly in the context of planetary 
nebulae, they can explain several of the observed features in this asymmetric 
H~{\sc i} structure. \citet{war07}, in particular, show that, in about 
$3\times10^4$ years into the post-AGB evolution, the central stellar system is 
not at the centre of planetary nebula and the nebula is highly asymmetric with 
the brighter emission in the downstream regions. For a density of 2 cm$^{-3}$ 
of the surrounding ISM, \citet{war07} find that atomic densities up to $50-100$ 
cm$^{-3}$ can be present in the ram pressure stripped flow in the downstream 
region. This is similar to and consistent with our estimated density of about 
10 cm$^{-3}$ in the H~{\sc i} structure for a distance of 13 kpc. 

Several AGB stars have been observed and detected in H~{\sc i} emission 
\citep{ger06,ger07} and many of these stars show a circumstellar envelope of 
size $\sim 1$ pc. More recently \citet{mat11} report the detection of a 
circumstellar wake and a compact high velocity cloud near the AGB star X 
Herculis. \citet{mat08} reported the detection of an H~{\sc i} tail coincident 
with part of the 4 pc long tail in the far ultraviolet trailing the AGB star 
o~Ceti. \citet{mar07} suggest that the long tail in this case is a result of 
the large space velocity ($\sim 130$ km~s$^{-1}$) of the Mira variable (which 
is a binary system) and the interaction between the ISM and the stellar wind. 
\citet{war07} concluded from their simulations of the system, that the long 
tail is a result of mass loss from the AGB star over $4.5\times10^5$ yr. 

Thus, we infer that (i) large H~{\sc i} structures and tails associated with 
stellar systems in the AGB or post-AGB phase are fairly common, and (ii) the 
observed asymmetry of the H~{\sc i} structure associated with V458~Vul may be 
due to the motion of the nova system with respect to the ambient ISM \citep[see 
also][for evidence of this in GK~Per]{bod04}. Unfortunately, such structures 
are difficult to detect due to the presence of Galactic emission kinematically 
corrupting the circumstellar emission and the inherently weak signal which 
limits detections only to nearby systems with peculiar velocities.

\subsection{Distance, age, and kinematics}
\label{sec:dak}

Distance estimates to novae have always been difficult to obtain and V458~Vul 
is no exception. From the different estimates in literature, it is clear that 
the distance estimate to V458~Vul is uncertain by at least a factor of two. 
\citet{wes08} estimate a range of 10 to 13.5 kpc from MMRD relation. Their 
estimations of the distance, based both on the light travel time argument and 
on the assumption that mean radial velocity of $V_{\rm LSR} = -60.6$ 
km~s$^{-1}$ is due to Galactic rotation, are also consistent with a distance 
of $\sim 13$ kpc. \citet{pog08} estimate the distance to be in the range 6.7 
to 10.3 kpc using several methods, some of them similar to \citet{wes08}. 
Recently, \citet{raj12} have reported a distance of $9.9 - 11.4$ kpc by 
modelling the expansion of ejecta based on optical interferometric 
observation. On the other hand, \citet{gor11} estimates the distance to be 5 
to 6.5 kpc using the method of light echoes \citep[although light echo 
modelling by][gives $d \sim 13$ kpc]{hou12}. Note that the line of sight 
velocity may be anomalous (like the high velocity clouds), and the actual 
distance may be very different from the kinematic distance. Also, with no 
reported observation before August 8, both the maximum magnitude and the rate 
of decline is highly uncertain in this case. Moreover, multiple peaks in the 
declining part of the optical lightcurve makes defining $t_2$ and $t_3$, and 
hence using MMRD relationship, non-trivial in this case. It is also important 
to note here that, even with exact determination of peak apparent magnitude 
and rate of decline, the intrinsic scatter of the MMRD relation 
\citep[e.g.][]{del95,dow00} and the uncertainty of extinction can result in a 
factor of $\sim 2$ uncertainty in the derived distance. 

\begin{table}
 \caption{Derived physical parameters of the H~{\sc i} structure for different assumed distances. The columns in the table are (1) Distance to the nova (D), (2) Size of the H~{\sc i} shell (d$_{\rm HI}$), (3) H~{\sc i} mass, (4) Estimated 
dynamical age ($t_{\rm dynamical}$), (5) proper motion ($\mu$) corresponding to the observed separation of V458~Vul from the centre of the H~{\sc i} shell, and (6) Tangential velocity ($v_t$) corresponding to a proper motion of 20 mas~yr$^{-1}$. See \S\ref{sec:dak} for details.}
\begin{center}
\begin{tabular}{c|c|c|c|c|c}
\hline
D    & d$_{\rm HI}$ & M$_{\rm HI}$ & $t_{\rm dynamical}$ & $\mu$         & $v_t$ \\
kpc  & pc           & M$_\odot$    & years               & mas~yr$^{-1}$ & km~s$^{-1}$ \\
\hline
13.0 & 7.6 & 25.0~ & $3.7\times 10^5$ & 0.27 & 1200 \\
~6.5 & 3.8 & ~6.3~ & $1.8\times 10^5$ & 0.56 & ~600 \\
~5.0 & 3.0 & ~3.7~ & $1.4\times 10^5$ & 0.71 & ~474 \\
~1.0 & 0.6 & ~0.15 & $2.8\times 10^4$ & 3.57 & ~~95 \\
\hline
\end{tabular}
\end{center}
\label{table:obs1}
\end{table}

Obtaining a reliable distance estimate to the nova system is important in 
determining the physical parameters of the H~{\sc i} structure. In 
Table~\ref{table:obs1}, the derived parameters for the system are presented 
assuming different distances to the nova. The columns in Table~\ref{table:obs1} 
are (i) assumed distance, (ii) size and (iii) mass of the H~{\sc i} shell, 
(iv) dynamical age of the shell for an average radius of $60\arcsec$ assuming 
an expansion velocity of 10 km s$^{-1}$, (v) proper motion corresponding to 
the observed separation of V458~Vul from the centre of the H~{\sc i} shell, 
and (vi) tangential velocity corresponding to a proper motion of 20 
mas~yr$^{-1}$ (see more on this below). Note that even if the line width is 
$\sim 5$ km s$^{-1}$, as we do not know the inclination, we have used a 
typical value of 10 km s$^{-1}$ for expansion velocity. If the expansion 
velocity on the sky plane is the same as the line of sight expansion velocity, 
the age will be about factor of two higher. Interestingly, in spite of the 
significant spatial displacement between the H~{\sc i} structure and the 
planetary nebula, and the difference in their estimated ages, their radial 
velocities are reasonably close ($V_{\rm LSR} = -53.2$ and $-60.6$ km~s$^{-1}$ 
respectively).

The star USNO-B1.0~$1108-0460444$ has been identified as the progenitor of 
V458~Vul by \citet{hen07}. The proper motion of the corresponding star listed 
by NOMAD\footnote{The Naval Observatory Merged Astrometric Dataset available 
online at {\tt http://www.nofs.navy.mil/nomad/}.} is about $-12 \pm 5$ 
mas~yr$^{-1}$ in Right Ascension and $16 \pm 2$ mas~yr$^{-1}$ in Declination 
(with the highest value of the probability estimator assigned for the 
likelihood that the quoted proper motion is correct). This is roughly $20 \pm 
3.5$ mas~yr$^{-1}$ in the north-west direction. For a distance of 13 kpc, the 
proper motion translates to an unreasonably large value of tangential velocity 
of $260 \pm 44$ AU per year i.e. $1200 \pm 200$ km~s$^{-1}$ (see 
Table~\ref{table:obs1}). On the other hand, the required velocity, to displace 
the nova by the observed separation of $\sim 100\arcsec$ from the centre of 
the H~{\sc i} shell over this dynamical timescale, is only $\sim 17$ 
km~s$^{-1}$ or 0.27 mas~yr$^{-1}$ at 13 kpc (see Table~\ref{table:obs1}). This 
is the vector sum of the proper motion of the star and the ejecta (which may 
be misaligned; \citealt{ger11}). So, the proper motion of V458~Vul is expected 
to be smaller than this. Also, V458~Vul is centred to within $1\arcsec$ of the 
14000 years old planetary nebula \citep{wes08}, implying proper motion smaller 
than 0.07 mas~yr$^{-1}$. It is more likely that the NOMAD proper motion is 
spurious for such a faint object (e.g., due to issues with the older plates 
used to derive proper motion).

We also examined the possibility that USNO-B1.0~$1108-0460444$ is {\it not} 
the progenitor of V458~Vul. At a distance of $\sim 13$ kpc, such a progenitor 
must be fainter than $V=21$ ($M_V > 3.5$ for $A_V= 1.95$). Taking a 
conservative lower limit of $M_V = 10.5$, and using a typical stellar 
luminosity function \citep{kro02}, the average number density of stars with 
$3.5 < M_V < 10.5$ is $\sim 2.5\times10^{-2} {\rm pc}^{-3}$. Considering the 
position uncertainty to be $\sim 0\arcsec.6$, the volume, defined by a circle 
of $0\arcsec.6$ radius (on the plane of the sky centred at the nova position) 
and a radial distance of $13\pm2$ kpc (2 kpc centred at a distance of 13 kpc), 
is about $9~ {\rm pc}^{3}$. Within this volume, the average expected number of 
progenitor stars with the right $M_V$ is about 0.22 (0.62 for $1\arcsec$ radius). 
The associated Poisson distribution probabilities are 0.18 and 0.33 for $0\arcsec.6$ 
and $1\arcsec$ respectively. Clearly, the possibility of an alternative nova 
progenitor at a distance of $\sim 13$ kpc is small.

If the H~{\sc i} structure has a physical association with the nova, as 
implied by various pieces of evidence above, then an indirect argument in 
favour of a smaller distance to the nova is from the estimated mass of the 
H~{\sc i} shell. At a distance of 13 kpc, the enclosed H~{\sc i} mass is 
$\sim 25$ M$_\odot$ which is too large to be explained as swept-up mass. If the 
distance is $\lesssim 6.5$ kpc, the H~{\sc i} mass may originate from a 
massive super-AGB progenitor. Such progenitors evolve to either neutron stars 
or massive ONeMg WDs \citep{her05}. Interestingly, the massive ($>1$ M$_\odot$) 
WD in V458~Vul is consistent with this scenario. If the NOMAD proper motion, 
as argued above, is incorrect, based on parameter values in 
Table~\ref{table:obs1}, a distance $\lesssim 6.5$ kpc seems plausible. It is 
important to constrain the proper motion of the faint nova progenitor, and if 
possible find tracers from the nova outburst which would give independent 
distance estimates, to cross check this suggestion of a smaller distance to 
V458~Vul.

\subsection{Possible large scale H~{\sc i} structure}

The above discussion has been mainly restricted to the shell-like expanding 
structure northeast of V458 Vul. However high velocity H~{\sc i} has been seen 
over a much larger region around V458 Vul. There are discrete clouds scattered 
around the nova with the central few arcminutes being devoid of H~{\sc i} (see 
Figure~\ref{fig:fig-1}). One possibility is that there is another larger shell 
which traces the mass loss history at a still earlier evolutionary epoch of 
the nova binary system. This large shell (approximately $8\arcmin$), with 
longer axis almost parallel to the smaller H~{\sc i} shell we discussed above, 
also has the nova off-centred. Moreover, this large shell seems to show weak 
evidence of expansion with the north/north-eastern parts moving at $-49$ 
km~s$^{-1}$ and the southern parts being close to $-55$ km~s$^{-1}$. However 
this needs confirmatory observations especially since GMRT is expected to be 
missing flux on larger angular scales. No clear correlation is seen between 
the H~{\sc i} and the IRAS 60 and 100 $\mu$m dust emission, though the 
H~{\sc i} shell appears to be located close to the edge of a dust cloud. 

\section{Concluding remarks}
\label{sec:con}

From the above discussion, we conclude that we have detected an expanding 
shell of cold H~{\sc i} possibly associated with V458~Vul. If this is the 
case, it appears to be the remnant of the interaction of the nova system with 
the ISM in its AGB phase of evolution, as indicated by its slowing down and 
its peculiar morphology. However, at a distance of 13 kpc, the H~{\sc i} 
structure would have an unreasonably high mass of 25 M$_\odot$. There are some 
indications that the nova is closer than 13 kpc, and further independent 
constraints on the distance to the nova will be very useful. 

Finally, it is relevant to mention that a large number of such systems might 
be distributed in the Galaxy. While we have detected the H~{\sc i} shell 
associated with V458~Vul, it is likely that there are several such structures 
which have not been detected in H~{\sc i} emission either because of the 
inadequate spectral resolution of such observations, or because they are so 
far from the central stellar system that the association is not obvious. More 
observations targeting such systems may be useful to understand the interaction 
with the surrounding ISM and the mass loss phase of the evolution of classical 
novae progenitor system.

\section*{Acknowledgements}
We thank the staff of the GMRT who have made these observations possible. GMRT 
is run by the National Centre for Radio Astrophysics of the Tata Institute of 
Fundamental Research. We are also grateful to the referee Anita Richards for a 
very useful review and for prompting us into substantially improving this 
paper. NR thanks M. Rupen and S. Bhatnagar for many helpful comments, and NGK 
thanks Rajaram Nityanada for discussions on proper motion.

\bsp

\label{lastpage}

\end{document}